\documentclass[12pt,a4paper]{iopart}

\usepackage{graphicx}
\usepackage{color}
\usepackage{amssymb}

\begin{document}
\title{Solution of semi-flexible self-avoiding trails on a Husimi lattice built with squares}
\author{Tiago J. Oliveira$^1$, W. G. Dantas$^2$, Thomas Prellberg$^3$ and J\"urgen F. Stilck$^4$}
\address{$^1$Departamento de F\'isica, Universidade Federal de Vi\c	
  cosa, 36570-900, Vi\c cosa, MG, Brazil} 
\address{$^2$Departamento de Ci\^encias Exatas, Universidade Federal
Fluminense, Volta Redonda, RJ 27255-125, Brazil}
\address{$^3$School of Mathematical Sciences, Queen Mary 
University of London, London E1 4NS, UK}
\address{$^4$Instituto de F\'{\i}sica, Universidade Federal
  Fluminense, Av. Litor\^anea s/n, 24210-346, Niter\'oi, RJ, Brazil} 
\ead{tiago@ufv.br, wgdantas@id.uff.br, t.prellberg@qmul.ac.uk, jstilck@id.uff.br}

\date{\today}

\begin{abstract}
  We study a model of semi-flexible self-avoiding trails, where the lattice paths are constrained to visit each lattice edge at most once, with configurations weighted by the number of collisions, crossings and bends, on a Husimi lattice built with squares. We find a rich phase diagram with five phases: a non-polymerised phase ({\bf NP}), low density ({\bf P1}) and high density ({\bf P2}) polymerised phases, and, for sufficiently large stiffness, two additional anisotropic ({\em nematic}) ({\bf AN1} and {\bf AN2}) polymerised phases within the {\bf P1} phase. Moreover, the {\bf AN1} phase which shows a broken symmetry with a preferential direction, is separated from the {\bf P1} phase by the other nematic {\bf AN2} phase. Although this scenario is similar to what was found in our previous calculation on the Bethe lattice, where the {\bf AN-P1} transition was discontinuous and critical, the presence of the additional nematic phase between them introduces a qualitative difference. Other details of the phase diagram are that a line of tri-critical points may separate the {\bf P1}-{\bf P2} transition surface into a continuous and a discontinuous portion, and that the same may happen at the {\bf NP}-{\bf P1} transition surface, details of which depend on whether crossings are allowed or forbidden. A critical end-point line is also found in the phase diagram.
\end{abstract}

\pacs{05.50.+q,05.70.Fh,64.70.km}

\maketitle

\section{Introduction}
\label{intro}
In poor solvents, polymer chains may undergo a transition from extended (coil) to collapsed (globule) configurations as the temperature is changed; this transition was called $\Theta$-point by Flory \cite{f66}. A grand-canonical experimental realisation of this collapse transition is seen in equilibrium polymerisation of sulphur solutions \cite{ks84}, and a lattice model where the interactions between the molecules in the solution, as well as the polymerisation process of sulphur rings are considered, reproduces well the experimental phase diagrams \cite{wp81}. In a certain limit \cite{ss92}, this model maps into a simpler one, where the effect of the solvent on the chain configurations (self-avoiding walks) is an attractive interaction between monomers in first-neighbour sites of the lattice which are not consecutive along a chain. This model became known as self-attracting self-avoiding walks (SASAW's), and has become the standard lattice model to study the collapse transition of polymers.   

It is convenient to display the thermodynamic behaviour of the grand-canonical model of SASAW's in the monomer-fugacity\,$\times$\,temperature plane. Two phases appear: a non-polymerised phase, which is just the empty lattice, and a polymerised phase, where a polymer chain occupies a fraction of the lattice sites with monomers. At high temperatures, when the Boltzmann factor associated with the attractive interactions is large, the polymerisation transition is continuous, so that the non-polymerised phase and a zero-density (extended) polymerised phase meet at the critical fugacity. As the temperature is lowered, the transition becomes discontinuous, so that the nature of the $\Theta$-point is tri-critical. This was found by de Gennes by mapping the polymer model into a ferromagnetic model \cite{dg75,dg79}. As the upper tri-critical dimension is three, mean-field tri-critical exponents, with logarithmic corrections, are found in three dimensions. This model has been solved on the two-dimensional hexagonal lattice by Duplantier and Saleur (DS) \cite{ds87} and, as expected, non-classical tri-critical exponents were obtained. As the DS solution requires some fine tuning of the model, it was discussed in the literature if it may be seen as the generic result for the collapse transition of polymers in two dimensions \cite{ss88,ds89,po94,gh95}, and these initial discussions seemed to furnish an affirmative answer to this question. This scenario may change if other phases are present in the phase diagram. For instance, on a Husimi tree built with squares if the interactions are supposed to be between bonds of the polymer chains located on opposite sides of an elementary square of the lattice, a {\em dense} polymerised phase, in which all lattice sites are occupied by monomers, is stable in part of the parameter space \cite{sms96}. The presence of this additional polymerised phase may change the nature of the collapse transition, where the continuous polymerisation transition line ends. This additional dense phase is present also in the model with bond-bond interactions on the square lattice \cite{mos01}.

In another lattice model which may describe the collapse transition the chains are represented by {\em trails}, where the excluded volume constraint is applied to lattice edges, that is, the walks which represent the chains are allowed to pass through each edge of the lattice at most once, so that they may visit each site up to $\left \lfloor{q/2}\right \rfloor$ times, where $q$ is the coordination number of the lattice \cite {mm75}. One notices that all SAW's are also trails, but if $q \ge 4$ there are trails which visit sites more than once, so that there are more trails than SAW's. The attractive interactions in the trails may now be on the same site, which is of course a simplification compared to the SASAW's model. On two-dimensional lattices, when more than two bonds are incident on a site, one should distinguish {\em crossings} from {\em collisions}. When crossings are allowed, the lattice is not planar anymore. If both crossings and collisions have the same statistical weight, we will call the model interacting SAT's (ISAT's). When the lattice is planar, that is, if only collisions are allowed, the model was called VISAW's (vertex-interacting SAW's) and solved exactly by Bl\"ote and Nienhuis (BN) \cite{bn89}. As the tri-critical exponents found in this solution are distinct from the ones in the DS solution, again there has been much discussion in the literature about which would be the generic model for the 2D collapse transition of polymers. The BN exponents seem difficult to find in simulations \cite{b13}, and if the chains are not longer flexible in the BN model, so that a bending energy is associated to elementary bends in the trails, both BN and DS tri-critical points are found in the phase diagram, united by a line of multi-critical points \cite{v15}. Recently, it has also been shown that the introduction of crossing in the VISAW's model is relevant, changing the universality class with respect to the BN case, where crossings are not allowed \cite{n16}.

In this paper, we study the general model of trails, with collisions, crossings and stiffness, on a Husimi lattice built with squares. In a previous work, we have studied the same model on a four-coordinated Bethe lattice \cite{d17}. The exact solution of models on such a hierarchical lattice may be considered an approximation of the behaviour of them on a regular lattice with the same coordination number. This is the reason why the thermodynamic behaviour of a model in the core of a Cayley tree is called its Bethe lattice solution, since it will often be identical to the Bethe approximation on a regular lattice \cite{b82}. As mean-field exponents are found in these solutions, they are not suited to study the point mentioned above regarding the generic universality class of the collapse transition. However, they may furnish phase diagrams which are qualitatively correct, and such information is sometimes difficult to obtain with more elaborate analytic techniques or simulations. A rich phase diagram was found solving the general trails model on the Bethe lattice, with regular ({\bf P}) and dense ({\bf DP}) polymerized phases, besides the non-polymerized phase ({\bf NP}). As in the solution of the ISAT model (where crossings and collisions have the same statistical weight) \cite{os16}, the continuous {\bf P}-{\bf DP} and {\bf NP}-{\bf P} transitions meet the discontinuous {\bf NP}-{\bf DP} transition. Thus, the collapse transition corresponds to a bi-critical point. However, for sufficiently stiff trails, an additional dense anisotropic ({\em nematic}) ({\bf AN}) polymerized phase is stable in a region inside the {\bf P} phase. The {\bf P}-{\bf AN} transition is discontinuous and critical, and this is a quite unusual transition, although it has been found in other models before and is understood in the framework of the normalisation group \cite{fb82}. In the limit of rigid rods, only the {\bf NP} and the dense isotropic ({\bf DP}) and anisotropic ({\bf AN}) phases remain and all transitions are discontinuous. The three coexistence lines meet at a triple point, which is the end-point of the bi-critical line. 

\begin{figure}[!t]
\centering
\includegraphics[width=7.0cm]{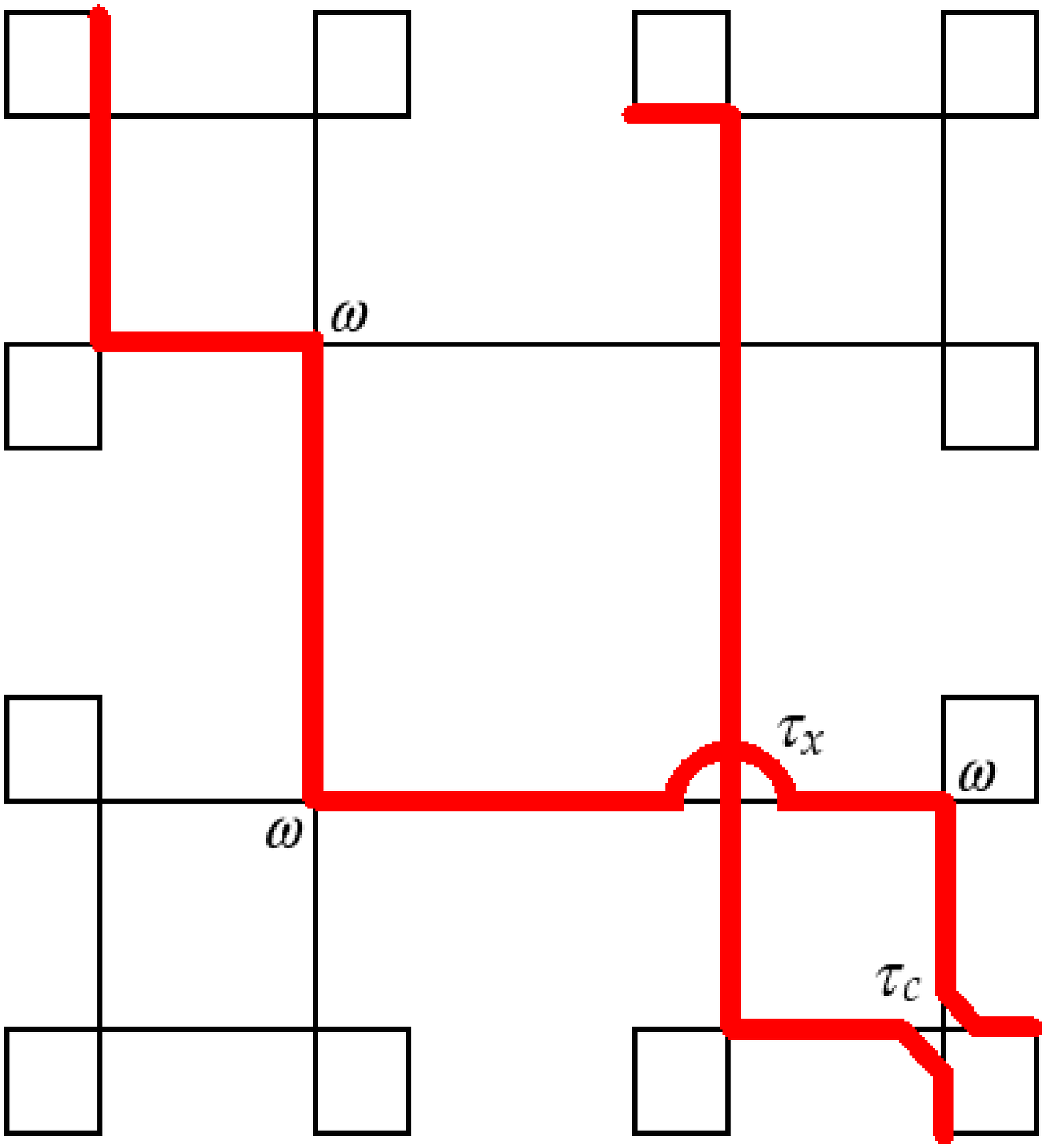}
\caption{(Colour online) Example of a contribution to the partition function of the model on a Husimi lattice with 3 generations. The statistical weight of this configuration is $z^{14} \tau_x \tau_c \omega^6$.}
\label{Figrede}
\end{figure}

Loops are not allowed in trails and SAW's, models with loops belong to other universality classes. This constraint poses a drawback of the Bethe lattice in the study of these models, since no closed loops are present on Cayley trees. For trails, this means in particular that collisions and crossings may be interchanged freely, which is not true on regular lattices. Thus, in the Bethe lattice solution the statistical weights of crossings ($\tau_x$) and collisions ($\tau_c$) always appear in the combination $\tau=\tau_x+2\tau_c$, so that it becomes interesting to study trails on lattices with loops. One of the simplest options is a Husimi lattice built with squares. Therefore, below we solve the general model for semi-flexible trails on a Husimi lattice with coordination $q=4$ built with squares. This lattice has a hierarchical structure, just as the Bethe lattice, thus allowing a solution to be obtained in terms of recursion relations for partial partition functions of sub-trees, with fixed root configuration. The trails have the end-point monomers placed on the surface of the tree. In figure \ref{Figrede} we have two trails on a Husimi tree with three generations of squares. The activity of each monomer is $z$ (as there are no endpoints of chains in the lattice, $z$ may also be seen as the activity of a bond), $\omega$ is the Boltzmann weight of an elementary bend in the trail, $\tau_x$ and $\tau_c$ are the Boltzmann factors of crossings and collisions, respectively.

Although many features of the phase diagrams for the model in the Bethe lattice solution \cite{d17} are found also here, there are significant qualitative differences. Among others, in the present solution we find {\em two} nematic phases ({\bf AN1} and {\bf AN2}), instead of the single one found before({\bf AN}), both of which are not dense. A consequence of this is that the discontinuous critical transition between the regular polymerized phase ({\bf P1}) and the nematic phase ({\bf AN}) for the Bethe lattice is replaced by two continuous transitions. Also, the transitions between the phases {\bf NP} and {\bf P}, as well as those between the phases {\bf P} and {\bf DP}, which are always continuous on the Bethe lattice, are now discontinuous in part of the interface between these phases, so that tri-critical lines are present in the phase diagrams. This leads to a change in the nature of the collapse transition, which was bi-critical before and now may be a tri-critical point or a critical end-point. 

The grand-canonical solution of the general model of trails on the Husimi lattice is presented in section \ref{shl}. The thermodynamical properties of the model are described in section \ref{tp}, the discussion of the phase diagrams in the canonical ensemble may be found in section \ref{can}, and final discussions and a conclusion are presented in section \ref{conc}.

\section{Solution on the Husimi lattice}
\label{shl}

As usual, to solve the model on the Husimi tree, we define partial partition functions (ppf's) for rooted sub-trees. They are called partial since the root configuration is fixed, and they are defined by the model. We then proceed writing down recursion relations for the ppf of a sub-tree with an additional generation connecting three sub-trees to a new root square. Iteration of these recursion relations furnishes ppf's of larger and larger sub-trees, so that the thermodynamic limit is reached after an infinite number of iterations. Finally, considering the operation of attaching four sub-trees to the central square of the tree, the partition function of the whole tree is obtained. All end-points of trails are placed on the surface of the tree.

\begin{figure}[!t]
\centering
\includegraphics[width=8.5cm]{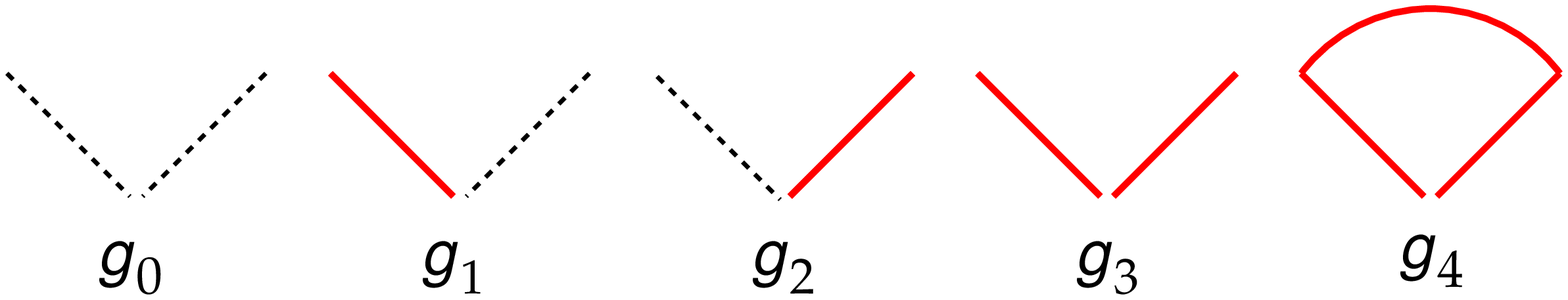}
\caption{(Colour online) Root configurations associated to the partial partition functions of the model on the Husimi lattice.}
\label{Figppfs}
\end{figure}

For a $q=4$ Husimi lattice built with squares (see Fig.~\ref{Figrede}), considering the number of incoming bonds in the root site and the constraint  of not having closed loops in the configurations, five ppf's are necessary, whose root configurations are depicted in Fig.~\ref{Figppfs}.  Note that in $g_4$ the two incoming bonds belong to the same chain, while in $g_3$ they are part of distinct chains, starting at different sites on the surface of the tree. We will make a distinction between the root configurations 1 and 2, where the polymer bonds are in different directions (horizontal or vertical in Fig.~\ref{Figrede}), since nematic ordering of the bonds may occur.

The recursion relations for the four ppf's are:
\numparts
\begin{eqnarray}
g'_{0} &=& a^3 + z a (b_1^2 + b_2^2) + z^2 b_1 b_2 c, \\
g'_{1} &=& z a^2 b_1 + z^2 a b_2 c + z^2 b_1^3 + z^3 b_1 c^2, \\
g'_{2} &=& z a^2 b_2 + z^2 a b_1 c + z^2 b_2^3 + z^3 b_2 c^2, \\
g'_{3} &=& z^2 a b_1 b_2 + z^3 (b_1^2+b_2^2) c + z^4 (c^3 - d^3),\\
g'_{4} &=& z^4 d^3,
\end{eqnarray}
where
\begin{eqnarray}
a &=& g_0 + \omega g_3,\\
b_1 &=& g_{1}+\omega g_{2}, \\
b_2 &=& g_{2}+\omega g_{1}, \\
c &=& \omega g_0 + (2\tau_c+\tau_x) g_3 + (\tau_c +\tau_x) g_4. \\
d &=& \omega g_0+\tau_c g_3+(\tau_c+\tau_x)g_4
\end{eqnarray}
\label{Eqppfs}
\endnumparts

The ppf's usually diverge upon iteration, so that we define ratios of them which may remain finite at the fixed point, associated to the thermodynamic limit. There are five possible definitions for the ratios, corresponding to the ppf which is placed in the denominator. If all root configurations survive at the fixed point, all choices are essentially equivalent, but it may happen that one or more of the ppf's vanish when compared to the others, and if the ppf of this particular configuration is placed in the denominator, ratios may diverge at the fixed point corresponding to these phases. We thus define the ratios $R_{i,k}=g_{i}/g_{k}$, and from the recursion relations for the ppf's we obtain recursion relations for them as
\numparts
\begin{eqnarray}
R'_{0,k} &=& \frac{A_{k}^3 + z A_{k} (B_{1,k}^2 + B_{2,k}^2) + z^2 B_{1,k} B_{2,k} C_{k}}{E_{k}},\\
R'_{1,k} &=& \frac{z A_{k}^2 B_{1,k} + z^2 A_{k} B_{2,k} C_{k}  + z^2 B_{1,k}^3 + z^3 B_{1,k} C_{k}^2}{E_{k}},\nonumber\\ \\
R'_{2,k} &=& \frac{z A_{k}^2 B_{2,k} + z^2 A_{k} B_{1,k} C_{k}  + z^2 B_{2,k}^3 + z^3 B_{2,k} C_{k}^2}{E_{k}},\nonumber\\ \\
R'_{3,k} &=& \frac{z^2 A_{k} B_{1,k} B_{2,k} + z^3 (B_{1,k}^2 + B_{2,k}^2) C_{k} + z^4 F_k}{E_{k}},\nonumber\\ \\
R'_{4,k} &=& \frac{z^4 D_{k}^3}{E_{k}},
\label{rrhus}
\end{eqnarray}
where $F_k=C_{k}^3-D_{k}^3$, and $E_{k}$ is given by the numerator of $R'_{k,k}$, so that $R'_{k,k}=R_{k,k}=1$ and only four RR's have to be iterated to find the stable phases of the system. In the expressions above
\begin{eqnarray}
A_{k} &=& R_{0,k} + \omega R_{3,k},\\
B_{1,k} &=& R_{1,k}+\omega R_{2,k},\\
B_{2,k} &=& R_{2,k}+\omega R_{1,k},\\
C_{k} &=& \omega R_{0,k} + (2\tau_c + \tau_x) R_{3,k} + (\tau_c + \tau_x) R_{4,k},\\
D_{k} &=& \omega R_{0,k} +\tau_c R_{3,k}+(\tau_c+\tau_x)R_{4,k}.
\label{param}
\end{eqnarray}
\endnumparts

The thermodynamic properties of the model are obtained in the limit when the number of iterations diverges, so that the stable phases are given by the stable fixed points of the RR's. This corresponds to the region in the parameter space ($z,\tau_x,\tau_c,\omega$) where the largest eigenvalue $\lambda_1$ of the Jacobian matrix evaluated at the fixed point:
\begin{equation}
J_{i,j;k}=\left( \frac{\partial R'_{i,k}}{\partial R_{j,k}}\right)_{R'=R},
\label{jac}
\end{equation}
is smaller or equal to one. The condition $\lambda_1=1$ defines the stability limit of a given phase.

The grand-canonical partition function of the model on the whole Husimi tree may be obtained considering the operation of attaching four sub-trees to the central square of the lattice, in a similar way as we used to derive the recursion relations for the ppf's. It can be conveniently written as $Y_k = g_{k}^{4} y_k$, with
\begin{eqnarray}
\label{Eqpfhusimi}
&&y_k = A_k^4 + 2 z A_k^2 (B_{1,k}^2 + B_{2,k}^2) + z^2 (B_{1,k}^4 + B_{2,k}^4) + \\ 
&& 4 z^2 A_k B_{1,k} B_{2,k} C_k + 2 z^3 (B_{1,k}^2 + B_{2,k}^2) C_k^2 + z^4 (C_k^4 - D_k^4), \nonumber
\end{eqnarray}
though it is actually independent of the choice $k$ for the denominator of the ratios of ppf's, as expected.

The density per \textit{site} of bonds, bends (in sites with a single monomer, visited only one time by the trail), collisions and crossings at the central square are, respectively: 
\begin{eqnarray}
\rho_z &=& \frac{z}{4 y_k}\frac{\partial y_k}{\partial z}\\
\rho_\omega&=& \frac{\omega}{4 y_k}\frac{\partial y_k}{\partial \omega}\\
\rho_c&=& \frac{\tau_c}{4 y_k}\frac{\partial y_k}{\partial \tau_c}\\
\rho_x&=& \frac{\tau_x}{4 y_k}\frac{\partial y_k}{\partial \tau_x}.
\end{eqnarray}

Note that the total density of bends is $\rho_{bends}=\rho_{\omega} + 2\rho_c$, since there are two bends at each collision. In addition, we define the densities of {\em single} incoming bonds in the root square
\begin{equation}
\rho_i = \frac{R_{i,k}}{4 y_k}\frac{\partial y_k}{\partial R_{i,k}},\\
\end{equation}
in directions $i=1$ and $i=2$. Note that $\rho_1$ and $\rho_2$ will be different (equal) in anisotropic (isotropic) phases, so that $Q=|\rho_1-\rho_2|$ is a convenient order parameter for the nematic phases. Another definition of the densities of bonds in each direction would be
\begin{equation}
\rho_i=\frac{\sum_j n_{j,i} a_{j,k}}{2y_k},
\end{equation}
where $a_{j,k}$ are the contributions to the grand-canonical partition function (\ref{Eqpfhusimi}) and $n_{j,i}$ is the number of bonds in direction $i$ at the central square for contribution $j$. It is possible to show that, as expected, at the fixed point both definitions lead to the same result, since the densities should be constant in the central region of the tree.  

The free energy of the model on the Husimi lattice, which is the core of the Husimi tree, can be found following the prescription proposed by Gujrati \cite{g95}. Considering each square of the Husimi lattice as a single site and connecting the adjacent sites by bonds, we obtain a Bethe lattice with coordination $q=4$, whose free energy per \textit{site} was calculated in detail in Ref. \cite{tiago16}. Therefore, the same result is obtained for the free energy per \textit{square} (divided by $k_BT$) on the Husimi lattice, being
\begin{equation}
 \phi_{b} = -\frac{1}{2} \ln\left( \frac{Y'_k}{Y_{k}^{3}} \right) 
\end{equation}
in the thermodynamic limit. Thus, from Eqs. (\ref{Eqppfs}) and (\ref{Eqpfhusimi}) one finds
\begin{equation}
 \phi_{b} = - \ln \left( \frac{E_k^2}{y_k} \right).
\label{Eqfe}
\end{equation}

\section{Thermodynamic properties of the model}
\label{tp}

\subsection{Results for $\omega=0$ (rigid rods)}

Before discussing the general case, we will study the limit where bends are forbidden ($\omega=0$ and, therefore $\tau_c=0$), so that we have rigid rods spanning the whole lattice. In this limit, both in the solution on the Bethe lattice and on the square lattice the same phase diagram was found \cite{d17}, with three phases: non-polymerized ({\bf NP}), anisotropic nematic ({\bf AN}), and dense polymerized ({\bf DP}). A similar phase diagram should be expected for the model on the Husimi tree. We start noting that for $\omega=0$, the recursion relations will be Eqs. (\ref{rrhus}), but with Eqs. (\ref{param}) replaced by:
\numparts
\begin{eqnarray}
A_k &=& R_{0,k},\\
B_{1,k} &=& R_{1,k},\\
B_{2,k} &=& R_{2,k},\\
C_k &=& \tau_x (R_{3,k} + R_{4,k}),\\
D_k &=& \tau_x R_{4,k}.
\label{paramw0}
\end{eqnarray}
\endnumparts
We notice that the root configuration $4$ will be present on the lattice only if we allow for it in the initial conditions, that is, on the squares of the surface of the tree. But even there they would imply only elementary bends of the chains, so that we should have $R_{4,k}=0$ for $k\neq 4$, when no bends are allowed. A fixed point $R_{1,0}=R_{2,0}=R_{3,0}=0$ is found here, which is identified with the {\bf NP} phase. The stability limit of this phase is determined by the eigenvalues of the Jacobian [Eq. (\ref{jac})], which is diagonal and has two eigenvalues equal to $z$, while the third vanishes. Thus, the {\bf NP} phase will be stable for $z \le 1$.

In the {\bf AN} phase, all edges in one of the two possible directions are occupied, and both directions are equally eligible. Supposing that the direction 1 is chosen, we consider ratios with $g_1$ in the denominator. The {\bf AN} fixed point will be at the origin in these ratios, and again the Jacobian is diagonal. The eigenvalues are $1/z$, $z\tau_x$ and $0$. Therefore, this phase is stable if $z \ge 1$ and $\tau_x \le 1/z$. 

Finally, in the {\bf DP} phase all edges of the lattice are occupied, so it is convenient to place the ppf $g_3$ in the denominator, so that this fixed point will be located at the origin. Besides a vanishing eigenvalue of the Jacobian, the other two are equal to $1/z\tau_x$, so this phase is stable if $\tau_x \ge 1/z$. 

\begin{figure}[t]
\centering
\includegraphics[width=8.3cm]{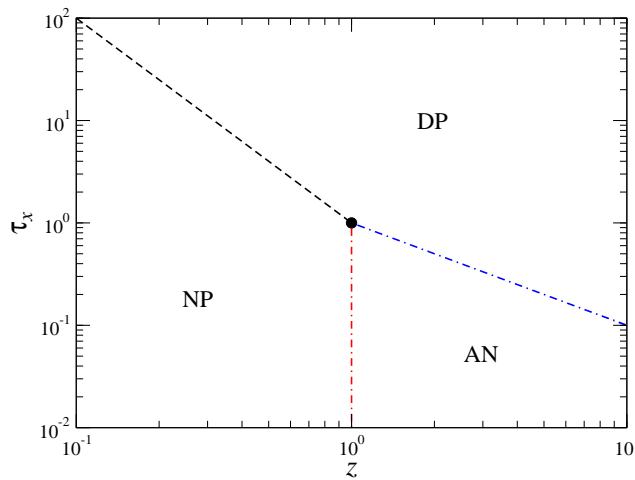}
\caption{Phase diagram for $\omega=0.0$ (rigid rods). The dashed and dash-dotted lines represent discontinuous and critical discontinuous transitions, respectively. \label{pd}}
\end{figure}

In summary, the {\bf AN} phase is separated from the {\bf NP} phase by a discontinuous critical transition at $z=1$, and from the {\bf DP} phase by a similar transition at $\tau_x=1/z$. These rather unusual type of phase transition was discussed in general by Fisher and Berker some time ago \cite{fb82}, and was also found and discussed in some detail in the solution of the present model on the Bethe and square lattices \cite{d17}. The {\bf NP}-{\bf DP} transition is also discontinuous, but not critical. It is located between the stability limits of both phases, and the free energies of both phases are equal at coexistence. Since $\phi_{b}^{(NP)}=0$ and $\phi_{b}^{(DP)}=-2\ln(z^2\tau_x)$, the coexistence line is given by $\tau_x=1/z^2$. The two discontinuous critical lines and the coexistence line meet at a bi-critical point in  $z=\tau_x=1$. As mentioned above, this phase diagram is identical to the ones found in the solution of this model on the Bethe and square lattices, and it is depicted in Fig.~\ref{pd} (figure 3d in \cite{d17}).

\subsection{Results for $\omega>0$}

When bends are allowed in the chains, we still have a non-polymerized phase (\textbf{NP}), similarly to rigid rods. The dense polymerized ({\bf DP}) phase [where $\rho_z = 1$ and $(\rho_c + \rho_x) = 1$] found in the solution of the model on the Bethe lattice \cite{d17}, gives place to a phase where most of the sites are occupied by two monomers, so that $\rho_z \approx 1$ and $(\rho_c + \rho_x) \approx 1$, but which is not dense anymore. Following Ref. \cite{tj16}, we will refer to it as the {\bf P2} phase. Furthermore, a second polymerized phase is stable for $\omega>0$, for which densities change with the statistical weights. This regular polymerized phase will be called {\bf P1}. There are now {\em two} nematic phases, the {\bf AN1} phase, which becomes the {\bf AN} phase when $\omega \to 0$, and a new nematic phase which we will denote by {\bf AN2} and will discuss below. The fixed points, considering $g_0$ in the denominator, are: 

\textit{i}) {\bf NP} phase: $R_{1,0}=R_{2,0}=R_{3,0}=0$ and $R_{4,0} > 0$, which leads to $\rho_i=0$, for $i=z,\omega,c,x$, and $\phi_b^{(NP)}=0$.

\textit{ii}) {\bf P1} phase: $R_{1,0}=R_{2,0}>0$, $R_{3,0}>0$ and $R_{4,0}>0$, with the densities $\rho_i$, for $i=z,\omega,c,x$, assuming values in the range $[0,1]$.

\textit{iii}) {\bf P2} phase: $R_{1,0}=R_{2,0}=0$, $R_{3,0}>0$ and $R_{4,0}>0$, where $\rho_z \approx 1$ and $(\rho_c + \rho_x) \approx 1$.

\textit{iv}) {\bf AN1} phase: $R_{1,0} \rightarrow \infty$ and $R_{2,0} \rightarrow \infty$, with $r \equiv R_{1,0}/ R_{2,0} $ assuming one of the values
\begin{equation}
r_{1,2} = \frac{1-3 \omega^2 \pm (1-\omega^2)\sqrt{1-4 \omega^2}}{2\omega^3}.
\label{eqrazao}
\end{equation}
Note that necessarily $r_1r_2=1$, since this phases exists only for $\omega<1/2$. We thus have a phase in which symmetry is broken, and there is nematic ordering. Moreover, in this phase $R_{4,0}=0$ and
\begin{equation}
R_{3,0} = \frac{z^2 \tau -1+\sqrt{W}}{2 \omega (1 + z \tau)},
\label{fp iv}
\end{equation}
where, $\tau \equiv \tau_x + 2 \tau_c$ and
\begin{equation}
W = z^4 {\tau}^2 + z^2 \tau [4 \omega^2 (z+1) - 2] + 4 z \omega^2 (z+1) + 1.
\end{equation}
Actually, for this phase it is more convenient to define the RR's with $g_2$ placed in the denominator, leading to the fixed point $R_{0,2}=R_{3,2}=R_{4,2}=0$ and $R_{1,2}=r$. In any case, we find $\rho_z=1/2$, $\rho_{\omega}=2\omega^4/[(1-\omega^2)(1-2\omega^2)]$ and $\rho_c=\rho_x=0$ for this {\bf AN1} phase, showing that all sites are occupied by a single monomer. Note that the densities turn out to depend only on $\omega$ in this phase. Considering the contributions to the partition function [Eq. \ref{Eqpfhusimi}], we note that the third term is dominant in this phase, so that most of the elementary squares of the tree are in configurations with two parallel bonds (on opposite edges), most of them in one of the two possible directions (1 or 2). Indeed, in this phase one finds 
\begin{equation}
\rho_{1,2}=\frac12\left(1\pm\frac{\sqrt{1-4\omega^2}}{1-2\omega^2}\right).
\label{density12}
\end{equation}
As expected, $\rho_1+\rho_2=1$, but $\rho_1\neq \rho_{2}$, and so the nematic order parameter 
\begin{equation}
Q=\frac{\sqrt{1-4\omega^2}}{1-2\omega^2}
\label{order}
\end{equation}
is non-zero, in opposition to $\rho_1=\rho_2$ (and $Q=0$) for the \textit{homogeneous} ({\bf NP}, {\bf P1} and {\bf P2}) phases. 

\begin{figure}[!t]
\centering
\includegraphics[width=8.3cm]{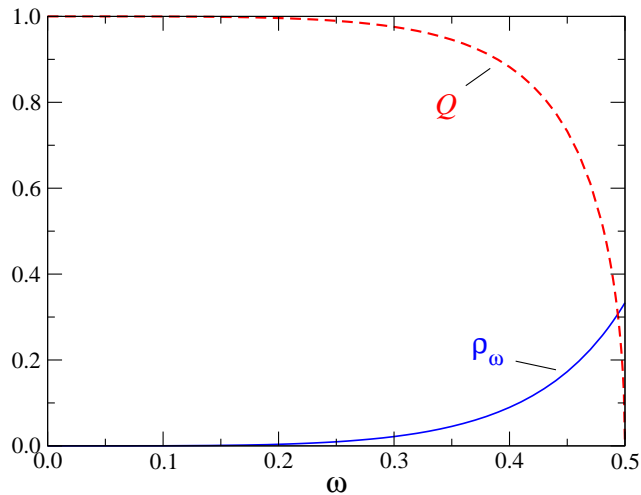}
\caption{Nematic order parameter $Q$ and density of sites with single bends $\rho_\omega$ in {\bf AN1} phase as functions of the Boltzmann factor for single bends $\omega$.}
\label{FigDens12Husimi}
\end{figure}

The order parameter $Q$ is displayed in Fig. \ref{FigDens12Husimi}, where $\rho_{\omega}$ is also shown. For high stiffness ($\omega \lesssim 0.2$), we see that $\rho_{\omega} \approx 0$ and $Q \approx 1$ so that the chains are almost all perfectly aligned in one direction (1 or 2). We notice that a similar phase was found for the sISAW model on the Husimi lattice \cite{p02}. Finally, the free energy of the {\bf AN1} phase can be easily calculated (considering the RR's for $R_{i,2}$), leading to
\begin{equation}
\phi_b^{(AN1)}=-\ln\left[ \frac{z^2 (1-\omega^2)^2}{1-2\omega^2}\right].
\label{feAN1}
\end{equation}
which also only depends on $\omega$.

\textit{v}) {\bf AN2} phase: $R_{1,0}> 0$, $R_{2,0}>0$, $R_{3,0}>0$ and $R_{4,0}>0$, with $R_1 \neq R_2$, so that there is again nematic order. The densities and the nematic order parameter in this phase interpolate between the values found in the phases between which it is stable, namely, the {\bf P1} and {\bf AN1} phases. Similarly to the \textbf{AN1} phase, this second nematic phase exists only for $\omega < 1/2$.

The {\bf NP} phase is stable in the region:
\begin{equation}
 (\tau_c+\tau_x) \leqslant \frac{S-1-2 z \omega}{z\left( 2 z + \frac{S-3}{\omega+1}\right) },
\label{eqLimNP}
\end{equation}
where
\begin{equation}
 S \equiv \sqrt{\frac{4-3 z (\omega+1)}{z (\omega+1)}}.
\end{equation}
For the {\bf AN1} phase, the stability region is given by
\begin{equation}
\tau \leq \frac{z(1-4\omega^2)-1}{z(z-1)},
\label{eqLimAN}
\end{equation}
with $\omega < 1/2$ and $z > 1/(1-4\omega^2)$. The stability limits of the {\bf P1}, {\bf P2} and {\bf AN2} phases were calculated numerically.

In the region of the parameter space between the stability limit of the {\bf AN1} phase (eq. \ref{eqLimAN}) and the one of the {\bf P1} phase (determined numerically), the nematic {\bf AN2} phase is stable, and its stability limits coincide with the ones of the two other phases, so that the {\bf AN1-AN2} and {\bf P1-AN2} phase transitions are continuous. As will be shown in the results presented below, the region of stability of the {\bf AN2} phase in the phase diagram is usually very small, so that at first view there is not much difference between the results on the Husimi lattice and the ones on the Bethe lattice. However, on the Husimi lattice the discontinuous critical {\bf AN1-P1} transition (found on the Bethe lattice) is replaced by two continuous transitions which are very close to each other, and so all densities and order parameters change continuously but rather fast in the region between these transitions.

We recall that the {\bf AN1} and {\bf AN2}  phases are stable only for $\omega<1/2$, whilst the other phases are stable for any $\omega$. Hence, the phase diagrams will present transitions between 3 and 5 phases for $\omega>1/2$ and $\omega<1/2$, respectively. In the following sections, we discuss in detail the results for the ISAT and the VISAW models.

\subsection{Results for the ISAT model}

First, we investigate the case where $\tau_x=\tau^*$ and $\tau_c=\omega^2 \tau^*$, so that crossings and collisions differ only through the stiffness present in the latter, leading to a semi-flexible ISAT (sISAT) model.

\subsubsection{Flexible case ($\omega=1$)}

For sake of completeness, let us start recalling the results for the classical (flexible) ISAT model, whose thermodynamic behaviour on a Husimi lattice built with squares and coordination $q=4$ was recently investigated by some of us \cite{tj16}. The phase diagram for this model presents critical (\textbf{NP-P1} and \textbf{P1-P2}) transition lines which meet a \textbf{NP-P2} coexistence line at a bi-critical point, located at $z=1/3$ and $\tau^*=\tau_x=\tau_c=3$. The phase diagram is shown in figure 12 of the reference above.

\subsubsection{Semi-flexible case ($0 < \omega < 1$)}

\begin{figure}[!t]
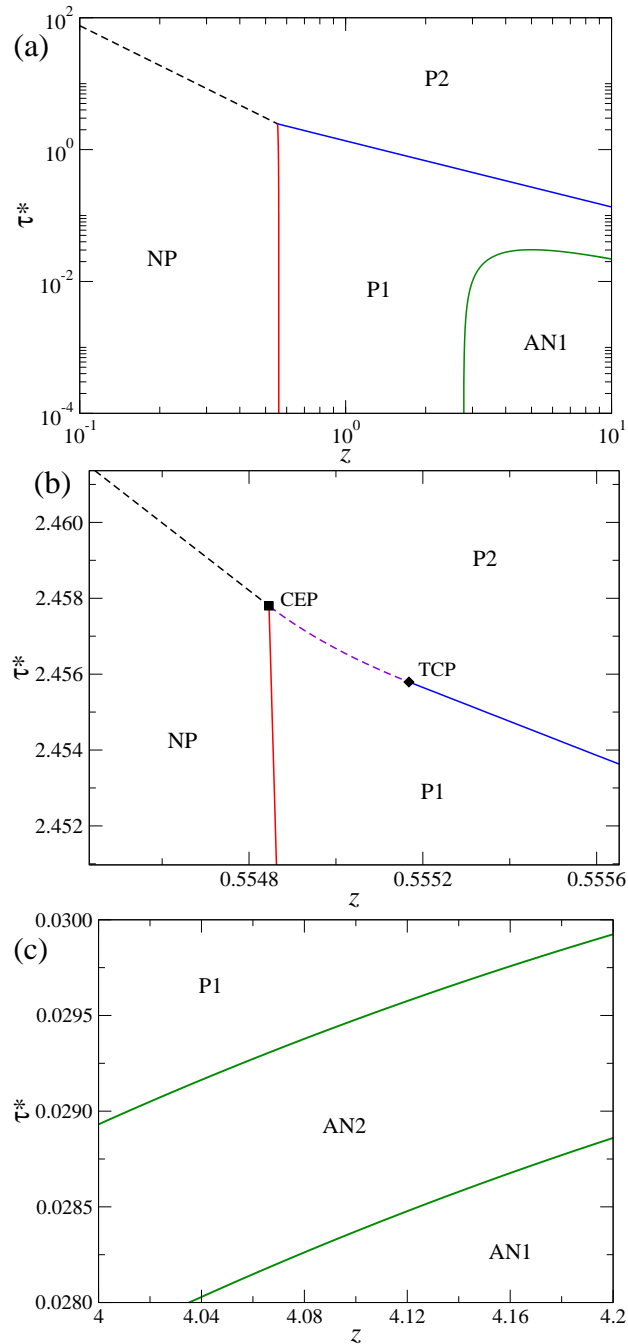

\centering
\includegraphics[width=8.1cm]{F5a.eps}
\includegraphics[width=8.1cm]{F5b.eps}
\includegraphics[width=8.1cm]{F5c.eps}
\caption{(Colour online) (a) Phase diagram for the sISAT with $\omega=0.4$ in the parameters $\tau^*$ versus $z$. The continuous transitions between {\bf NP}-{\bf P1} (red) and {\bf P1}-{\bf P2} (blue) phases are given by continuous lines. The dashed lines indicate {\bf NP}-{\bf P2} (black), and {\bf P1}-{\bf P2} (violet) coexistence lines. The last one is visible only in panel (b), which highlights the region around the critical end-point (CEP - square) and the tri-critical point (TCP - diamond). The continuous (green) line denotes both continuous transitions between {\bf AN1}-{\bf AN2} and {\bf AN2}-{\bf P1} phases, which are not distinguishable in the graph (a), but may be seen in an enlarged picture of this region in panel (c).}
\label{FigHusimiw040ISAT}
\end{figure}

When stiffness is included in the system, instead of a bi-critical point, as found for $\omega=1$, one observes a tri-critical point (TCP) and a critical end-point (CEP) in the phase diagrams. See an example in Fig. \ref{FigHusimiw040ISAT}, for $\omega=0.40$, where one finds: 
\begin{itemize}
 \item A continuous {\bf NP}-{\bf P1} transition line, which ends at the CEP;

 \item A discontinuous {\bf NP}-{\bf P2} transition line, which also ends at the CEP;

 \item Continuous and discontinuous {\bf P1}-{\bf P2} transition lines, which meet at the TCP. The discontinuous transition starts at the CEP and ends at the TCP, where it gives place to the continuous {\bf P1}-{\bf P2} line;
 
 \item Continuous {\bf AN1}-{\bf AN2} and {\bf AN2}-{\bf P1} transition lines.
 
\end{itemize}

The two transitions involving nematic phases are quite close, so they in general are are not distinguishable in a scale showing all features of the phase diagram, the same happens with the tri-critical point and the critical end-point. For any $\omega$ in the range $(0,0.5)$, one finds phase diagrams with the same properties shown in Fig. \ref{FigHusimiw040ISAT}. Hence, the general 3D phase diagram (in variables $z,\tau^*,\omega$) presents four ({\bf P1}-{\bf NP}, {\bf P1}-{\bf P2}, {\bf AN1}-{\bf AN2}, and {\bf AN2}-{\bf P1}) critical surfaces, two ({\bf NP}-{\bf P2} and {\bf P1}-{\bf P2}) coexistence surfaces, a line of critical end-points (CEP line) and a tri-critical line (TC line). The only difference for $\omega \in [0.5,1)$ is the absence of the nematic phases {\bf AN1} and {\bf AN2} and, consequently, of the {\bf AN1}-{\bf AN2}, and {\bf AN2}-{\bf P1} critical surfaces. In general, the {\bf NP}-{\bf P1} critical and the {\bf NP}-{\bf P2} and {\bf P1}-{\bf P2} coexistence surfaces meet at the CEP line, and the {\bf P1}-{\bf P2} critical and coexistence surfaces meet at the TC line.

\begin{figure}[!t]
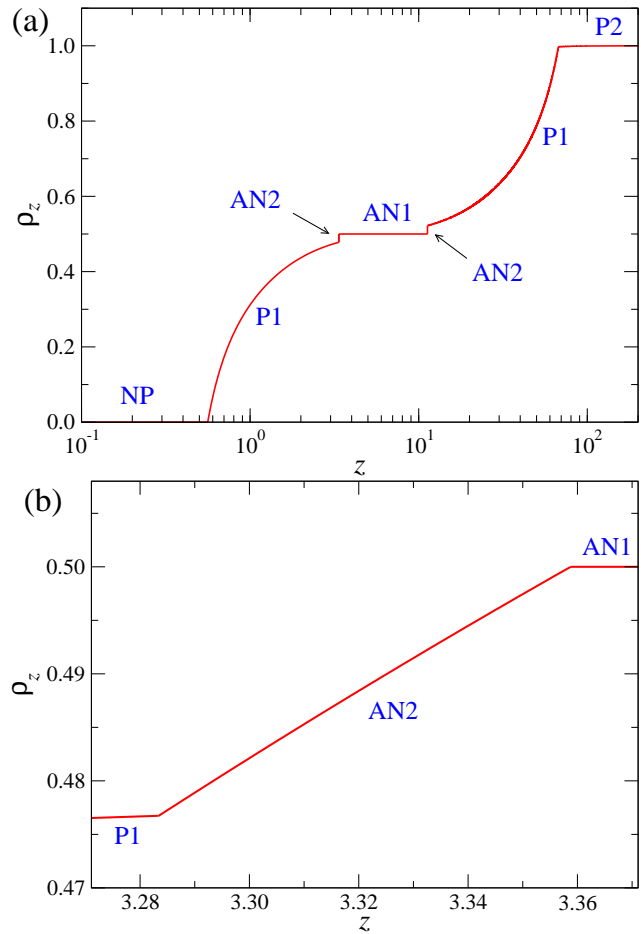

\centering
\includegraphics[width=8.3cm]{F6a.eps}
\includegraphics[width=8.3cm]{F6b.eps}
\caption{(Colour online) (a) The density of bonds for sISAT with $\omega=0.4$ and $\tau^*=0.02$ as a function of $z$. The value of $\tau^*$ is chosen such that all four phases are accessed upon varying $z$. (b) An enlarged view of the region around the {\bf P1}-{\bf AN2}-{\bf AN1} transitions.}
\label{density}
\end{figure}

To further elucidate this scenario, Fig. \ref{density} shows the density of occupied sites at $\omega=0.4$ and $\tau^*=0.02$ as $z$ increases. At this value of $\tau^*$, the system passes through all five phases as $z$ increases, from the {\bf NP} phase, in which the density is zero to the {\bf P2} phase in which the density is nearly equal to one. Note in particular that within the {\bf P1} phase the {\bf AN1} phase appears in which the density of occupied sites is equal to $0.5$. Note also in panel (b) that between the {\bf P1} and {\bf AN1} phases the density changes continuously, since the system passes through the {\bf AN2} phase. For this value of $\omega$, within the {\bf AN1} phase, the density of vertical and horizontal bonds equals $0.4706$ and $0.294$, and the density of bends equals $0.0896$.

\begin{figure}[!t]
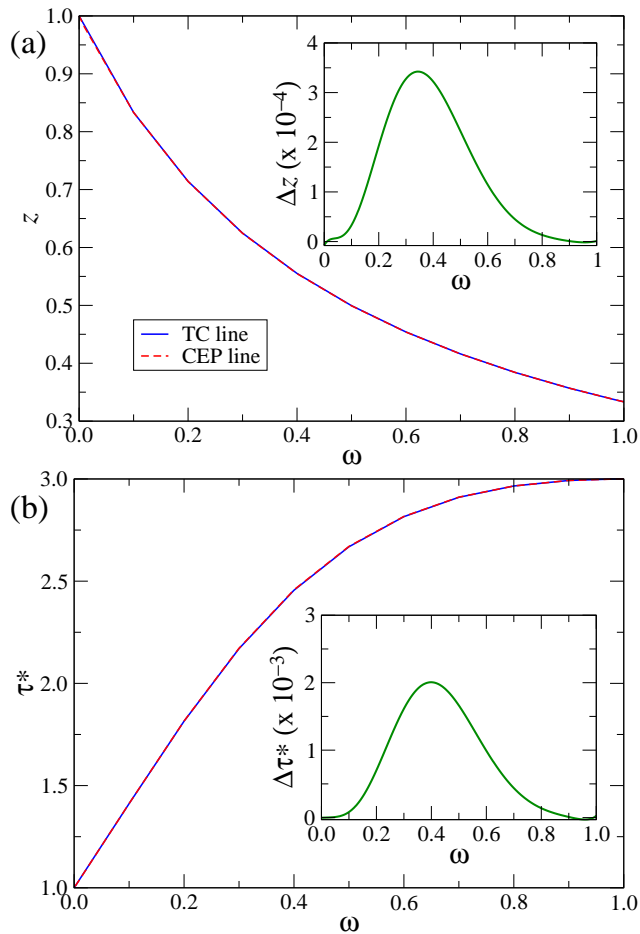

\centering
\includegraphics[width=8.3cm]{F7a.eps}
\includegraphics[width=8.3cm]{F7b.eps}
\caption{(Colour online) Variation with $\omega$ of the coordinates (a) $z$ and (b) $\tau^*$ of the tri-critical and critical end-point lines, for the sISAT model. The differences between the coordinates of both lines as function of $\omega$ are shown in the insets. In (a) $\Delta z \equiv z_{TC}-z_{CEP}$ and in (b) $\Delta \tau \equiv \tau_{CEP}-\tau_{TC}$.}
\label{FigCEPTCPHusimiISAT}
\end{figure}

Figure \ref{FigCEPTCPHusimiISAT} shows the differences between the coordinates ($\tau^*$ and $z$) of the CEP and TCP as functions of $\omega$. As one can see, these lines are always very close and the differences between them (shown in the insets) seem to vanish in the limits $\omega \rightarrow 1$ and $\omega \rightarrow 0$. This suggests that the CEP and TC lines meet at the bi-critical point when $\omega \rightarrow 1$, showing that this point is actually a multi-critical point. This multi-critical behaviour is consistent with recent results by Pretti \cite{p16}, who has analysed the flexible model ($\omega=1$), but considering general $\tau_c \neq \tau_x$. When $\omega \rightarrow 0$, the CEP and TC lines also seem to meet at a single multi-critical point, which turns out to be the point where the transition lines (critical discontinuous {\bf NP}-{\bf AN} and {\bf DP}-{\bf AN} and coexistence {\bf NP}-{\bf DP}) meet in phase diagram for rigid rods ($\omega=0$), discussed in the previous section. As will be demonstrated in the following subsection, again, this point looks like a bi-critical point in the phase diagram for $\omega=0$.

\subsection{Results for the VISAW model}

Now, we turn our attention to the case where $\tau_x=0$, with $\tau_c=\omega^2 \tau^* \neq 0$, so that crossings are forbidden, corresponding to a semi-flexible VISAW (sVISAW) model.

\begin{figure}[!t]
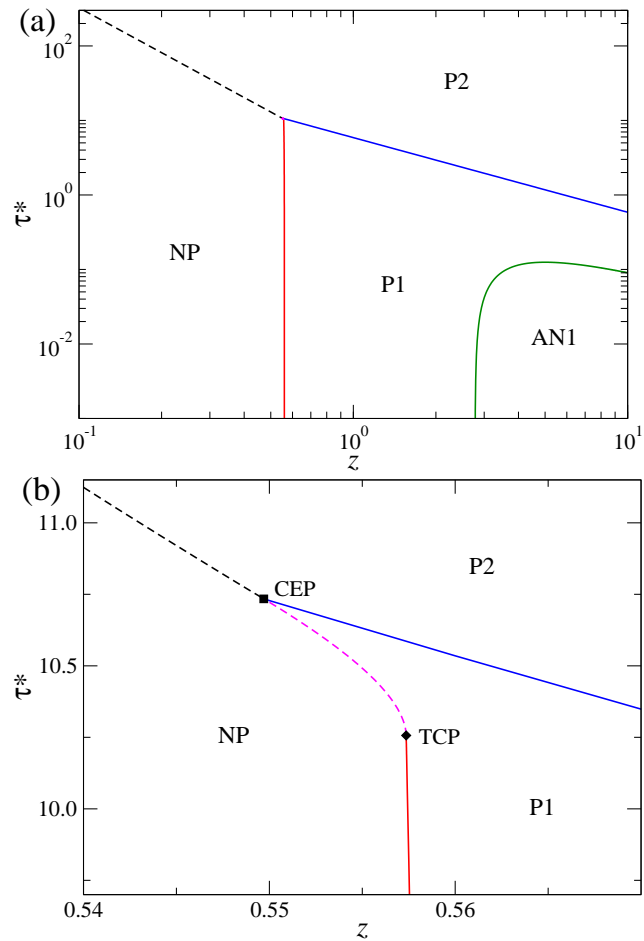

\centering
\includegraphics[width=8.3cm]{F8a.eps}
\includegraphics[width=8.3cm]{F8b.eps}
\caption{(Colour online) (a) Phase diagram for the sVISAW with $\omega=0.40$. The continuous transitions between {\bf NP}-{\bf P1} (red) and {\bf P1}-{\bf P2} (blue) phases are indicated by continuous lines. The continuous (green) line indicates the two continuous transitions {\bf P1}-{\bf AN2} and {\bf AN2}-{\bf AN1}, indistinguishable in the scale of the plot. The dashed lines are the {\bf NP}-{\bf P2} (black) and {\bf NP}-{\bf P1} (magenta) coexistence loci. (b) Details of the region around the critical end-point (CEP - square) and the tri-critical point (TCP - diamond).}
\label{FigHusimiw040VISAW}
\end{figure}

The thermodynamic behaviour of the flexible ($\omega=1$) VISAW model on a ($q=4$) HL built with squares has been investigated by Pretti in \cite{p16}. Although this model presents transitions between the same phases of sISAT's, there are differences in their phase diagrams. Namely, there exists a {\bf NP}-{\bf P1} critical line for small $\tau^*$, but this transition becomes discontinuous at a tri-critical point. The {\bf NP}-{\bf P1} coexistence line ends at a critical end-point (CEP), where it meets a continuous {\bf P1}-{\bf P2} line (this transition is always continuous here) and a discontinuous {\bf NP}-{\bf P2} line. See Fig. \ref{FigHusimiw040VISAW} for an example of this behaviour.

\begin{figure}[!t]
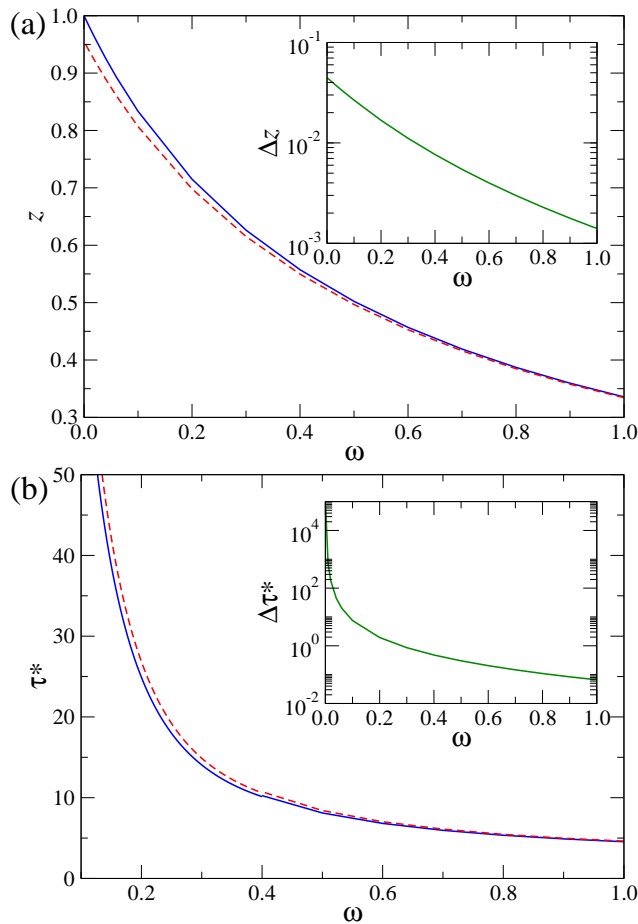

\centering
\includegraphics[width=8.3cm]{F9a.eps}
\includegraphics[width=8.3cm]{F9b.eps}
\caption{(Colour online) Variation with $\omega$ of the coordinates (a) $z$ and (b) $\tau^*$ of the tri-critical and critical end-point lines, for the sVISAW model. The differences between the coordinates of both lines as function of $\omega$ are shown in the insets. In (a) $\Delta z \equiv z_{TC}-z_{CEP}$ and in (b) $\Delta \tau \equiv \tau_{CEP}-\tau_{TC}$.}
\label{FigCEPTCPHusimiVISAW}
\end{figure}

For any $\omega$ in the range $(0,1]$ one still finds, qualitatively, the same transitions between the phases ({\bf NP}, {\bf P1}, and {\bf P2}). Furthermore, for $\omega < 1/2$ the {\bf AN1} and {\bf AN2} phases are also stable and, again, two continuous transition lines/surfaces separate the {\bf AN2} phase from the {\bf AN1} and {\bf P1} phases. As this aspect of the phase diagram is very similar to what was found for ISAT's, we do not show detailed results of these transitions here. See the phase diagram for $\omega=0.40$ in Fig. \ref{FigHusimiw040VISAW} for an example. Thence, the general phase diagram (in the variables $z,\tau^*,\omega$) presents four ({\bf NP}-{\bf P1}, {\bf P1}-{\bf P2}, {\bf P1}-{\bf AN2}, and {\bf AN1}-{\bf AN2}) critical surfaces, two ({\bf NP}-{\bf P1} and {\bf NP}-{\bf P2}) coexistence surfaces, a CEP line and a TC line. At the CEP line the {\bf NP}-{\bf P1} and {\bf NP}-{\bf P2} coexistence surfaces, as well as the critical {\bf P1}-{\bf P2} surface meet. At the TC line the {\bf NP}-{\bf P1} critical and coexistence surfaces meet. 

Interestingly, the overall phase diagrams of sISAT and sVISAW are similar, but there is an exchange: In sISAT (sVISAW) the critical {\bf NP}-{\bf P1} ({\bf P1}-{\bf P2}) surface ends at a CEP line and a TC line is associated with {\bf P1}-{\bf P2} ({\bf NP}-{\bf P1}) transitions. Another important difference is the bi-critical/multi-critical points in sISAT, which do not exist in sVISAW. Namely, here the CEP and TC lines do not meet at a single point when $\omega \rightarrow 0$ or $\omega \rightarrow 1$. We notice that the case $\omega=0$, which leads to $\tau_c = 0$, is trivial here, since the only non-null thermodynamic parameter is the fugacity $z$. So, there are just two continuous transitions in this case: {\bf NP}-{\bf P1} at $z=1$ and {\bf AN}-{\bf P1} at $z=2$.

Figure \ref{FigCEPTCPHusimiVISAW} shows the variation of the coordinates ($z$ and $\tau^*$) of the CEP and TC lines with $\omega$. In contrast with sISAT, here an appreciable difference exists between these curves, whose magnitude is shown in the insets of Fig. \ref{FigCEPTCPHusimiVISAW}. Another difference from sISAT, is the divergence in $\tau_{CEP}^*$ and $\tau_{TC}^*$ as $\omega \rightarrow 0$. We note that, for both lines, $\tau_c$ approach finite values in this limit and, so, by definition, $\tau^* = \tau_c/\omega^2$ diverges. We recall that such divergence (non-divergence) in sVISAW (sISAT) is consistent with the results for Bethe lattice (see figure 6 in Ref. \cite{d17}) and physically expected, as already discussed in that case.

\section{Mappings on the canonical ensemble}
\label{can}

In the canonical ensemble, for chains with a given size $N$, the key thermodynamic parameters of the semi-flexible ISAT and VISAW models are the energies $\varepsilon$ and $\varepsilon_b$ associated, respectively, with monomers in double visited sites and bends in the chains. Thereby, the energy of a chain with $N_b$ bends and $N_t$ collisions and crossings is $E = - N_t \varepsilon + N_b \varepsilon_b$, so that $\tau = e^{\beta \varepsilon}$ and $\omega = e^{-\beta \varepsilon_b}$, where $\beta=1/(k_B T)$, $k_B$ is Boltzmann's constant and $T$ the temperature. The canonical situation, or equivalently the single chain limit, corresponds to the critical and coexistence surfaces separating the {\bf NP} phase from the others. The critical {\bf NP}-{\bf P1} surface corresponds to swollen (coil) chains, the {\bf NP}-{\bf P1} coexistence (in sVISAW) to collapsed (globule) chains, and the {\bf NP}-{\bf P2} coexistence is associated with a more dense collapsed phase. 

\begin{figure}[!t]
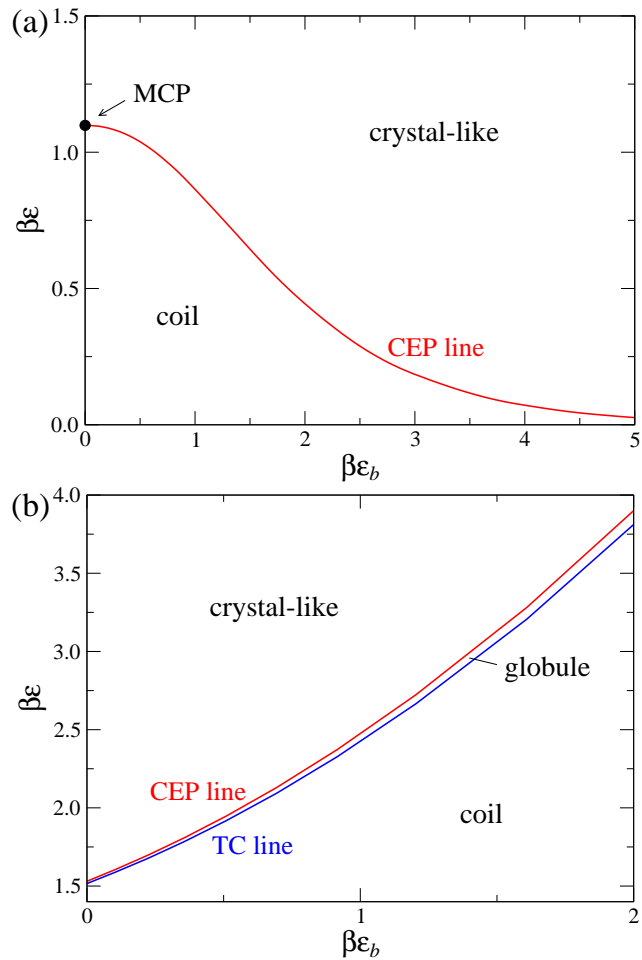

\centering
\includegraphics[width=8.3cm]{F10a.eps}
\includegraphics[width=8.3cm]{F10b.eps}
\caption{(Colour online) Canonical phase diagrams for the (a) sISAT and (b) sVISAW models, in variables $\beta \varepsilon = \ln \tau^*$ against $\beta \varepsilon_b = - \ln \omega$.}
\label{FigDiagCanonicoHusimi}
\end{figure}

In sISAT, the transition between the coil and both phases is given by the CEP line, for $0 < \omega < 1$, and by the multi-critical points at $\omega=0$ and $\omega=1$. This canonical phase diagram (in the variables $\beta \varepsilon_b \times \beta \varepsilon$) is shown in Fig. \ref{FigDiagCanonicoHusimi}a. This mapping suggests that different behaviours might be found in canonical investigations of the sISAT in flexible and semi-flexible cases. It is noteworthy that evidence of discontinuous transitions has been found in canonical Monte Carlo simulations of some similar models \cite{kpor06,doukas}, while CEP lines have been observed in their Bethe/Husimi solutions \cite{tiago,tj16}. This suggests that discontinuous transitions could be also found in canonical MC simulations of the sISAT model. Since in the grand canonical phase diagram the tri-critical line is very close to the critical end-point line for the Husimi lattice solution, a different scenario may appear on the square lattice, for example. We recall that on the Bethe lattice the collapse transition is bi-critical \cite{d17}.

Figure \ref{FigDiagCanonicoHusimi}b shows the canonical phase diagram for the sVISAW. In this case, a coil-globule transition associated with a TC line is observed, which is similar to the $\Theta$-point behaviour of the classical ISAW model. Notwithstanding, within our mean-field framework, we cannot assure that the sVISAW TC line is in the $\Theta$-universality class. Curiously, the globule phase is observed only in a tiny region of the parameter space, being limited from above by a CEP line, where the (canonical) chains undergoes a transition to a crystal-like phase. 

\section{Final discussions and conclusions}
\label{conc}

In this work we presented a model of self-avoiding trails, where the lattice paths are constrained to visit each lattice edge at most once, with configurations weighted by the number of collisions, crossings and bends, on a Husimi lattice built with squares. As in our previous study of such a model on a Bethe lattice \cite{d17}, we find rich phase diagrams with: A non-polymerized phase ({\bf NP}), low density ({\bf P1}) and high density ({\bf P2}) polymerized phases, and two anisotropic ({\em nematic}) ({\bf AN1} and {\bf AN2}) polymerized phases.

However, there are intriguing differences: While on the Bethe lattice the {\bf P2} phase is maximally dense, and the continuous {\bf P1}-{\bf P2} and {\bf NP}-{\bf P1} transitions meet the discontinuous {\bf NP}-{\bf P2} transition in a bi-critical point, we find that for semi-flexible interacting self-avoiding trails on the Husimi lattice the continuous {\bf P1}-{\bf P2} transition ends at a tri-critical point, and the bi-critical point is replaced by a critical end point, which is connected to the tri-critical point via a coexistence line between the {\bf P1} and {\bf P2} phases. In contradistinction, if crossings are forbidden, we find that the continuous {\bf NP}-{\bf P1} transition ends at a tri-critical point, which connects to a critical end point via a coexistence line between the {\bf NP} and {\bf P1} phases.

For sufficiently stiff trails, two additional anisotropic ({\em nematic}) ({\bf AN1} and {\bf AN2}) polymerized phases are stable in a region inside the {\bf P1} phase. While on the Bethe lattice the {\bf P1}-{\bf AN} transition is discontinuous and critical, we find here two continuous transitions, which separate the {\bf AN2} phase from the {\bf P1} and the {\bf AN1} phases. The {\bf AN1} phase is no longer totally ordered nematically, as it was on the Bethe lattice solution. Thus, the rather unusual discontinuous critical {\bf P1}-{\bf AN} transition found on the Bethe lattice is replaced by two continuous transitions, very close to each other in the parameter space. While the densities of collisions and cross-links vanish in the {\bf AN1} phase, this is not the case in the {\bf AN2} phase, in which all densities vary interpolating between the values at the two phases on its limit of stability. Since the solution on the Husimi lattice should be a better approximation to the behaviour of the model on regular lattices as the one on the Bethe lattice, one possibility would be that on the square lattice a single nematic phase, with properties similar to the {\bf AN2} phase, appears inside the {\bf P1} phase, with a continuous transition between them. The phase diagram for the sVISAW model presented in \cite{v15} is closer to the result we obtained on the Bethe lattice, since they suggest that two critical surfaces meet a coexistence surface, at what they call a multi-critical line, which resembles a bi-critical line. As seen above, the tri-critical line is very close to the CEP line we found on the HL. In general, we notice that the qualitative differences between the BL and HL solutions are all restricted to quite small regions of the parameter space of the model.

\section*{Acknowledgements}

We acknowledge the support of CNPq and FAPEMIG (Brazilian agencies), particularly from the former through project PVE 401228/2014-2, which made the stay of T.P. in Brazil possible. TP gratefully acknowledges support by EPSRC grant EP/L026708/1.

\end{document}